\documentstyle[11pt,aasms4]{article}

\def \ref {\noindent\hangindent=1.0in\hangafter=1}
\def \cl {\centerline}

\def\ltsima{$\; \buildrel < \over \sim \;$}
\def\simlt{\lower.5ex\hbox{\ltsima}} 
\def\gtsima{$\; \buildrel > \over \sim \;$}
\def\simgt{\lower.5ex\hbox{\gtsima}} 

\begin{document}

\title{ROSAT, ASCA AND OSSE OBSERVATIONS OF THE \\
       BROAD LINE RADIO GALAXY 3C120}

\author{P. Grandi\altaffilmark{1},  R. M. Sambruna\altaffilmark{2}, 
L. Maraschi\altaffilmark{3}, G. Matt\altaffilmark{4},}
\author{C. M. Urry\altaffilmark{5}, R. F.  Mushotzky\altaffilmark{2}}

\bigskip
\noindent
$^1$ Istituto di Astrofisica Spaziale - CNR, Via E. Fermi 21, I-00044
Frascati (Roma), Italy

\noindent 
$^2$ NASA/GSFC Code 662, Greenbelt, MD 20771

\noindent 
$^3$ Osservatorio Astronomico di Brera, via Brera 28, I-20121 
Milano, Italy

\noindent
$^4$ Dipartimento di Fisica, Universita' degli Studi "Roma Tre", Via
della Vasca Navale 84, I-00146 Roma, Italy

\noindent
$^5$ Space Telescope Science Institute, 3700 San Martin Drive,
Baltimore, MD 21218

\begin{abstract}
We present simultaneous observations of the superluminal
radio galaxy 3C~120 performed
with the ASCA and GRO (OSSE) satellites on February-March 1994, as well as an
analysis of all the ROSAT archival data.
The ASCA spectrum of this object
can be described by an absorbed ($N_H=1.6\times 10^{21} {\rm cm}^{-2}$)
power law with photon index $\Gamma_{ASCA}=2$
and a very broad $(\sigma> 0.8$~keV) intense
iron line (EW$>400$ eV) at $\sim 6$~keV.

The combined ASCA--OSSE data do not exclude the presence of a
narrower ($\sigma=0.4$ keV) and less intense (EW $< 300$ eV) iron line plus
a hard component, corresponding either to reflection from an accretion disk
or to a flatter power law from a jet.
However a single power law plus
broad Fe line is preferred from a statistical point of view by the ASCA data

The ROSAT data yield a column density in excess of the Galactic value.
The spectral slopes, ranging from $\Gamma_{ROSAT}=2.5$
to 3.3, are steeper than that measured by ASCA, suggesting
the presence of a soft excess.
The 0.1-2~keV power-law slope is variable and softer
at higher intensity.

These results show that the combined soft and hard X-ray spectrum
of 3C~120 is rather complex.
The intrinsic absorption, the soft excess,
and the iron line indicate that the X-ray emission from this blazar-like
radio galaxy is dominated by a Seyfert-like component,
at least in the 0.1-10~keV energy band.
The jet contribution, if present, becomes important only at higher energies.
\end{abstract}


\section{Introduction}
3C~120 is a relatively nearby (z=0.033) 
core-dominated broad-line radio galaxy with a
superluminal jet extending from 0.5~pc to 100~kpc and diffuse radio
structures in several directions (Walker et al. 1987). A faint
optical counterpart of the radio jet has recently been detected
in deep high-resolution images (Hjorth et al. 1995).

Although 3C~120 is classified as a Seyfert 1 galaxy, its optical morphology is
not simple. Photometric and spectroscopic studies of the weak
nebulosity around the nucleus indicate the presence of two components,
a spheroidal system plus an extended disk, which suggests 3C~120 may
have resulted from the merging of
two different galaxies (Moles et
al. 1988). The indication of a possible double nucleus in 
monochromatic emission line images (Hua 1988) and the observation of
extended, strongly perturbed emission line regions (Tadhunter et
al. 1989; Baum, Heckman, $\&$ van Breugel 1992) support this
hypothesis.
The origin of the photons ionizing the extended emission-line regions
is still debated; the photoionizing flux may derive 
from the nuclear AGN continuum, or from the interaction of the
jet with the interstellar medium, or from local young stars 
(Baldwin et al. 1980, Balick $\&$ Heckman 1979, Hua 1988, Moles et al. 1988)

At UV wavelengths, 3C~120 has a typical AGN spectrum with a strong blue
bump and strong emission lines (Maraschi et al. 1991, hereafter M91). 
In the X-ray band, a simple absorbed power law gave an adequate
representation of the EXOSAT spectrum (M91).
No soft excess was detected with 
EXOSAT, while earlier {\it Einstein} Solid State Spectrometer (SSS) 
data indicated the presence of
emission lines at low energies (Petre et al. 1984, Turner et al. 1991).
No information about the iron line at 6.4~keV or possible Compton
reflection from a disk was available before the ASCA observations
because 3C~120 was never observed with
Ginga. At higher energies, it has been detected several times 
in the energy band 50-150~keV with OSSE 
(Johnson et al. 1994), but never at MeV or higher energies
with Comptel or EGRET (Maisak et
al. 1995, von Montigny et al. 1995). 

The X-ray spectral variability of 3C~120 is unusual. The energy spectral
index in the 0.3-6~keV band, measured with EXOSAT, varied between $\alpha\sim 0.5$ and
$\alpha\sim 0.8-1$ and was positively correlated with the soft
luminosity, but uncorrelated with the hard luminosity (M91). A possible interpretation of this behavior is that the
beamed radiation from the jet contributes only to the hard X-ray emission. 
Seyferts generally become softer when the
X-ray luminosity increases, while blazars, dominated by jet emission, 
become harder as their intensity increases
(Giommi et al. 1989; Grandi et al. 1992, Sambruna et al. 1994).

The detection of thermal features in the X-ray spectrum of 3C~120 is a very
important diagnostic of the presence of nonthermal emission
from the jet. 
The featureless beamed X-ray continuum will tend
to dilute any spectral features, such as absorption edges or
emission lines often found in Seyfert galaxies,
when its  relative contribution is important.
In this regard, no conclusions could be derived up to now because of the lack of
data with sufficient sensitivity at both soft and hard X-ray energies. 

Motivated by the complexity of this radio galaxy, we
organized simultaneous observations of 3C~120
with ASCA and OSSE-GRO covering the energy range 0.5--150~keV. 
Archival ROSAT PSPC data of this source, covering the energy range 
0.05--2.0~keV, were also analyzed. With such broad energy coverage,
we have for the first time the possibility of strongly constraining
the X-ray emission mechanisms and understanding more about the 
innermost structure in 3C~120.

\section{ASCA Observations}

\subsection{Observations and Data Analysis} 

We observed 3C~120 with ASCA for 50,000~sec on 1994 February
17-19 (the journal of the observations is given in Table~1).
The two Solid-state Imaging Spectrometers (SIS0,SIS1) operated
in faint mode during the observation, with only one CCD used (1-CCD
mode) and the two Gas Imaging Spectrometers (GIS2,GIS3) in Pulse
Height mode. 
The average source flux during the pointing was $F_{2-6~{\rm keV}}
\sim 3\times10^{-11}$ ergs cm$^{-2}$ s$^{-1}$. 
Compared to the measured EXOSAT fluxes (M91), 
3C~120 was in an intermediate brightness state during our ASCA 
observation.

Data reduction for all 4 instruments on ASCA (SIS0, SIS1, GIS2, GIS3;
Tanaka, Inoue, $\&$ Holt 1994) followed the standard procedures
in the ABC Guide (Day et al. 1994).
Specifically, screening criteria include the rejection
of data taken during intervals of Earth and elevation 
angles lower than 20$^{\circ}$ and 10$^{\circ}$, 
respectively, and
magnetic rigidity lower than 6 GeV/c for the SIS detectors and 
10 GeV/c for the GIS detectors. 
Data taken during passage through the South Atlantic anomaly
were also rejected. 

The SIS and GIS spectra were accumulated on circular regions
centered on the source position with radii 4 arcmin and 6 arcmin,
respectively, due to the different point-spread functions of the two
instruments. 
The GIS background was calculated from the same source image, while
the SIS background from blank-sky
observations (available from the ASCA Guest Observer Facilities).
In the SIS case, we opted for blank sky observations because the 
brightness of the source and the broad PSF strongly reduced the regions
where the local background could be measured.
In both cases, the background spectrum was accumulated within regions
having the same dimensions as the source circle
and from blank-sky.

The spectra were rebinned in order to have at least 20 counts per
new bin, to validate the use of the $\chi^2$ statistic in the
spectral fits. 
The more recent versions (1994 November 9)
of the standard redistribution matrices (RMF) of the SIS and GIS were taken
from the ASCA Calibration Database on the public node legacy.gsfc.nasa.gov, 
while the ancillary files (ARF) have been appropriately built for
each instrument with the program `ascaarf' (version 2.62) of the 
FTOOLS package. 
Spectral fits were performed using the XSPEC package. 
 
For each instrument, light curves with different bin size (100, 500, 1000
sec) were extracted in XSELECT and analyzed using the XRONOS package.

\subsection{ Timing Analysis}

Significant time variability was detected in all the instruments,
in agreement with the findings of Nandra et al. (1996a) for 3C120.
A clear increasing trend is present in both the SISs and GISs, the flux
changing $\sim30\%$ during the observation (see Figures 1a-1b).
A standard $\chi^2$ test, applied to the average count 
rate in each
light curve, confirms the variability of 3C~120: the probability that 
the source was constant is less than $10^{-3}$ in both the SISs and GISs 
light curves, independent of the bin size used. The $\chi^2$ values are 
listed in Table~2 for each instrument and for each bin size.

\subsection{ Spectral Fits }

We performed spectral fits on data from the individual and combined instruments.
Since we did not find significant differences, we discuss here only the
results from the combined instruments (SIS0+SIS1, GIS2+GIS3, and 
SIS0+SIS1+GIS2+GIS3). 

The spectral fit parameters are reported in Table 3.
The listed uncertainties correspond to the
90\% confidence interval for one parameter of interest ($\Delta
\chi^2 = 2.7$). 
When the parameters are not well constrained by XSPEC, we report only
an upper limit. 

We first fitted the SIS data in a soft band, 0.6-5~keV,
to avoid the complexity in the Fe spectral region.
Because of the uncertain SIS calibration at low energies, 
we excluded the data below 0.6~keV.
A single absorbed power law with
$\Gamma=2.00\pm0.02$ and N$_H=1.64 (\pm0.07)\times10^{21}$ cm$^{-2}$ 
is sufficient to fit the SIS spectra very well (see table 3).
The fitted column density is consistent with Galactic
${N_H}^{\rm Gal}=1.23\times10^{21}$ cm$^{-2}$ (Elvis et al. 1989), 
especially allowing    
for the systematic SIS overestimate of N$_H= 2$-$3\times10^{20}$
cm$^{-2}$.

In order to investigate the hard X-ray emission of 3C~120, the spectral fits
were then extended to energies $>5$~keV. We fitted the SIS data from 0.6
to 8~keV and the GIS data from 1 to 10~keV.
A single absorbed power law gives an unacceptable $\chi^2$ both for
the two SISs and GISs (Table~3) with residuals showing an excess around 
6~keV (Fig. 2a-2b). When a Gaussian emission line
is included in the model, the $\chi^2$ improves significantly
(the F-test indicates the addition of two parameters represents
an improvement at a confidence level $>99\%$ for both the SIS and GIS data.)
We alternately added a Gaussian emission line to a power law
with fixed slope ($\Gamma=2$, as suggested by the soft spectrum) and
to a power law with free slope. The results are similar,
but obviously the uncertainties are smaller in the first case. The position
of the line ($E_{Fe}=6.06^{+0.26}_{-0.21}$ in SISs
and $E_{Fe}=6.30^{+0.20}_{-0.19}$ in GISs) is consistent, within 
the uncertainties, with 
fluorescent iron from cold matter 
(observed $E_{Fe_{K_\alpha}}$=6.4/1.033=6.196~keV).

The iron line is unambiguously detected at 99$\%$ confidence, as shown in 
Figure 3 where the confidence contours  of 
the line width versus  the line normalization are plotted for a power law 
with $\Gamma$ free to vary.
 
The observed intrinsic width of the line
($\sigma\sim 0.8-1 $~keV) and its equivalent width
(EW$> 400$ eV) are rather large. (The Gaussian fitted to the combined GIS 
spectra is
somewhat larger than that for the two SIS, probably because of the different energy
resolution and ranges covered by the instruments.)
We investigated different
possibilities for the line broadening. 
We first tested an emission
profile from a relativistic disk around a Schwarzschild black hole
(Fabian et al. 1989). We fitted the higher-resolution
SIS data with a power
law with fixed slope ($\Gamma=2$) illuminating an accretion disk with
inner and outer radii 
$r_i=10$ and $r_o=1000$ (in units of $GM/c^2$).
The inclination of the disk was fixed at 
$i=15^{\circ}$; this is the upper limit of the angle to the line of sight 
for the
radio jet, as 
deduced from its apparent superluminal velocity
$\beta_{app}\sim 8$ (Zensus 1989).
As shown in Table~3,  for the constrained inclination angle the $\chi_r^2$ 
is worse for the diskline model than the gaussian line model.
A low-inclination disk can not account
for the observed emission line. Relaxing the inclination angle, the
model fit requires a viewing angle close to 90$^{\circ}$, which is
difficult to reconcile with the superluminal properties of the
jet (if orthogonal to the disk)
and with the brightness of the blue bump.

The broad Gaussian profile of the iron line is also preferred to 
a blending of two
narrow lines. Two Gaussian profiles with narrow fixed intrinsic widths
($\sigma_1=\sigma_2=0.1$~keV) can not adequately reproduce the SIS
emission feature (see Table~3). When we allowed the widths to
vary, the fitted normalization of one of the two Gaussians converges
to zero.

We checked whether the addition of a hard component could change the
X-ray continuum shape and affect the iron line in both the SIS and
GIS data. Two different models were used, (i) a Compton reflection
and (ii) a broken power law. 
\noindent
A reflection model is directly suggested
by the presence of the $K_{\alpha}$ line (Lightman $\&$ White 1988,
Matt et al. 1991, George $\&$ Fabian 1991). 
A flatter second power law is
expected if the jet
contributes to the X-ray spectrum. 
We parameterized the reflection component using the PLREF model in
XSPEC. It assumes the illumination of a cold disk from an X-ray source
and uses the Lightman and White (1988) approximation to calculate the shape
of the reflected spectrum. 

(i) In order to better constrain the hard component, the power-law slope
was frozen at $\Gamma=2$ and all the input parameters, except the
ratio (R) between the normalization of the power law 
and the reflection component, were fixed to the default values. 

A Gaussian centered
at 6.2~keV and with intrinsic width 0.4~keV (a ``standard'' iron
line shape, as suggested by Nandra et al. 1996b) was added,
with parameters fixed at first and then free to
vary. As shown in Table~3, the hard component ($R=0.8$ and
$R=1.6$ in the SIS and in the GIS, respectively) contributed to
the continuum and reduced the EW of the line if the Gaussian profile
was defined ``a priori'', but it was not necessary anymore
($R<0.44$) if the iron line parameters were free to vary. 
In this latter case the line is again very broad.


This result reveals that the  amount of reflection and the 
line parameters are strongly correlated. This is clearly shown in Figure 4
where the GIS2+GIS3 probability contours of the line width versus ($\sigma$)
versus the ratio (R) between the normalization of the power law 
and the reflection component are plotted.

(ii) Similar results were obtained fitting the data with two power laws
plus a Gaussian. If the energy and the intrinsic width of the line
were frozen ($\sigma=0.4$~keV), the power law was significantly
flatter above the energy break (estimated around $\sim 4$~keV) and the
iron equivalent width was small. If we allowed the line to broaden, the
hard power-law slope became steeper (the photon index became
comparable to the value below the break) and the EW of the line
increased again.
Simultaneous fits to all four instruments confirm these
results (Table~3). 
In this case, we obtained only an upper limit for
the normalization of the reflected component if 
the iron line profile is gaussian.

We then conclude that it is not possible with the ASCA data alone
to constrain the contribution of a
hard component that, if present, could affect the measured strength and
profile of the line. 
The actual detection of an
emission feature at 6.2~keV, however, does ensure that the contribution
of the jet is small below $\sim8$~keV.

\section{OSSE Observations}

3C~120 was observed with OSSE from February 17 to March 1 1994,
overlapping with the ASCA observations. A shorter pointing 
was also performed a week later, from March 8 to March 15.
The source was very faint. In both the observations,
it was detected at only a 2$\sigma$ level of significance 
(c/s= 0.084$\pm0.042$ and c/s= 0.090$\pm$0.045 in the first 
and second pointing, respectively). 
Because of this, we could not obtain an OSSE
spectrum on February-March 1994, but only an average flux from the sum of both 
observations.

In order to constrain the shape of the high energy emission,
the more 
plausible models for the ASCA data 
were extrapolated to 150~keV and compared to the observed OSSE flux.
In particular we considered a simple power law plus a broad iron line,
and a power law plus a
hard component with a ``standard'' iron line shape.
As can be seen in Figure~5, none of the models can be definitely rejected.

Using non-contemporaneous OSSE fluxes from literature, the 
high energy spectrum appears to be well fit by a harder model.
In Figure~5, the open point represents the weighted 
average flux from all the 1992-1993 
detections of 3C~120 in the 50-150~keV band (Johnson et al. 1994).
In this case, however, the non-simultaneity of the ASCA and OSSE data
limits our ability to constrain the ASCA extrapolation.

\section{ ROSAT Observations}

3C~120 was observed with the ROSAT Positional Sensitive Proportional
Counter (PSPC) in pointed mode on four separate occasions in March,
September, and August 1993. The log of the observations is reported in
Table~1. We analyzed
the archival PSPC data using XSELECT. The events were corrected for
spatial gain changes using the ``pcsasscor'' tool (ROSAT Status
Report, No. 137, April 1, 1996). The data were extracted in a circle 
centered on the target position and with radius 2 arcmin, which is large 
enough to collect all the softest photons from the source and avoid the 
``ghost-image" problem. The background was estimated in a circle of
the same radius at $\sim$ 4 arcmin away from the source. According to the WGA 
catalog (White, Giommi, $\&$ Angelini 1994), any serendipitous sources in 
the field of 3C~120 are 
too faint 
($<$ 0.01 c/s) to affect our analysis.

Source spectra were extracted in 256 PI channels and rebinned on the
channel range 12-211 (energy range 0.15--2.1~keV) in order to have at
least 20 counts per new bin. 
Since the observations
were performed in low-gain mode (i.e., after October 1991), we used
the pspcb-gain2-256 matrix for the spectral fits, as appropriate.

\subsection{ Timing Analysis}

The source intensity varied among the four observing epochs, from a
minimum count rate in the energy range 0.1--2.1~keV of 1.74 $\pm$ 0.03
cts/s in March 1993 to a maximum count rate of 3.01 $\pm$ 0.04 cts/s in
August 1993 (Table~1). 

In order to test for variability of the flux within the single
exposures, we applied the Kolmogorov-Smirnov (KS) test on unbinned
data and the $\chi^2$ test on binned data. Flux variability was
detected for the 1993 September 10 observation, at 99\% confidence
according to the KS test. In order to apply the $\chi^2$ test we
rebinned the light curve on time scales $\ge$ 400 s, in order to avoid
spurious variability due to the satellite wobble. The light curve with
bins at 400 s is shown in Figure~6. Constant emission is ruled out at
99\% confidence.

\subsection{Spectral Fits }

The ROSAT data were first fitted separately with a single power law
with free N$_H$. The fit parameters 
are reported in Table~4  together with their
90\% confidence uncertainties for one parameter of interest. 

In all cases the fitted absorption is 
 significantly ($>$ 99\% confidence) higher than the
Galactic value, ${N_H}^{\rm Gal}$ = 1.23$ \times 10^{21}$ cm$^{-2}$.
This confirms
previous results based on {\it Einstein} IPC and SSS data (Kruper,
Urry, $\&$ Canizares 1990; Turner et al. 1991). 
The ROSAT data are characterized by a continuum with photon index
$\Gamma$=2.5--3.3, systematically steeper than that measured 
at higher energies by ASCA ($\Delta\Gamma>0.5$).

\noindent
There is
some evidence ($\sim$ 90\% confidence) for variation of the both
spectral parameters with source intensity. The column density
increases and the photon index gets steeper
from March 6, when the source was fainter, to the brighter
state of August 25. The steepening of the spectrum with increasing
the soft intensity is consistent with the EXOSAT results (M91).

The difference between ASCA and ROSAT spectral slopes suggests 
the presence of a soft excess at low energies.
 
It is known that calibration problems affect the PSPC results: 
Fiore et al (1994) showed a systematic difference between 
IPC-Einstein and ROSAT-PSPC spectra of $\Delta \Gamma\sim 0.2$.
However, we exclude that the discrepancies between ASCA and ROSAT results
are only due to calibration problems.
The ROSAT power laws are steeper than ASCA and always $\ge 0.5$.
In addition the ASCA slope is well determined. It is
measured over a large energy band
and appears in good agreement with the previous results 
from HEA0-1, {\it Einstein}-MPC 
and EXOSAT (Rothschild et al. 1983, Treves et al. 1988, M91).

In order to model the soft excess, we added a Raymond Smith model to 
the simple power law and fixed $N_H$ to the Galactic value 
(as suggested by ASCA data).
The results are listed in Table ~5. 
Acceptable $\chi^2$ are obtained for all the epochs, but the additional 
component does not significantly improve the fit. 
However, the spectral indices in Table ~5  are more close to 
the values usually observed at higher energies and the 
temperature (0.6-0.8 keV)
is in reasonable agreement with the result of Turner et al. (1991).
 
Summarizing, there is evidence in the ROSAT data of flux variability
on both long ($\sim$ months) and short ($\sim$ minutes)
time scales. 
There is evidence of a complex soft component.
Spectral variability is confirmed, with the spectral slope getting 
steeper when the intensity increases.

\section{Discussion and Conclusions}

3C~120 is a variable and complex X-ray source.
The Sefert-like spectrum is characterized by a steep power law ($\Gamma=2.0$), 
by an iron line from cold matter 
and by a probable soft excess.   
At higher energies (50-150 keV), our observations do not allow to 
strongly constrain the spectral shape. The OSSE measurements are consistent 
with a simple power law extrapolated by the ASCA data, but do not
exclude the presence of an additional hard component. \par

In details,  
the major observational results of this study are as follows:

\noindent
1. The ASCA data have revealed the presence of an iron line in 3C~120
for the first time. If the continuum 
is modeled with a single power law, the iron line is broader 
($\sigma\sim 0.9$~keV) and more intense ($\sim500$~keV) than in 
most Seyfert 1 galaxies. An object with 
a similar and intense broad line is the Seyfert 2 galaxy, 
IRAS 1832 (Iwasawa et al. 1996).

\noindent
2. A hard component (from a disk or jet) can fit the 
hard ASCA spectrum but is not required. 
However, the amount of reflection (or the jet contribution)
anti-correlates so strongly with the line parameters that it is difficult
to constrain their relative contributions to the spectrum.

\noindent
3. The February-March 1994 OSSE data are consistent with the ASCA results.
The 50-150~keV flux is very close to the extrapolation of the ASCA power law.
A reflected power-law component (or a second power law) predicts a slight excess of photons 
in the OSSE band but is consistent within the uncertainties.

\noindent
4. The ROSAT PSPC spectra are characterized by strong absorption 
and steep spectral indices.
We interpret this result as indication of soft excess. 

\noindent
5. Flux variability on short (hours) and long (months) time scales 
is observed
in the 0.1-2~keV energy band. Variations of $\sim 30\%$
are also seen by ASCA on time scale of days.
The short-time-scale variability is remarkable. While time variability 
on time scales of
days and months had already been
seen with the EXOSAT
satellite, this is the first time that fast variability on time scale of hours 
is observed in this source. The short time scale variability, notably
the factor 1.2 decrease in the 0.1-2~keV flux 
in $\sim 8000$~sec, indicates that the region producing the 
photons can not be larger than 
$\sim r_{\rm X-ray} < c t_D \sim 10^{15}$ cm, where the doubling
time scale, $t_D\equiv (F_{\rm i} / \Delta F) t_{\rm var}$,
is the time necessary for the source flux to change by a factor of two.

\noindent
6. There is an indication of spectral variability in the soft X-ray range. 
In particular, the 
spectral slope increases with intensity, as already noted in 3C~120 
(M91) and as is generally observed in 
Seyfert~1 galaxies (Grandi et al. 1992).\par

\bigskip

These results indicate that 
the X-ray emission in 3C~120 is similar to those in radio 
quite Seyfert galaxies.
The excess absorption of soft photons and
the soft excess suggested by the ROSAT spectra as well as 
the well-resolved ASCA iron line exclude
a strong contribution of beamed radiation between 0.1 and $\sim 8$~keV. 

3C~120 is not the only case of an intense radio source in which 
the iron line is 
seen. Emission features at 6.4~keV have also been detected with ASCA
in other radio galaxies, for example, 3C390.3 (Eracleous
et al. 1996),  3C445 (Yamashita $\&$ Inoue 1996) and 3C109 (Allen et al. 1997).
This suggests that the jet intensity in the 0.6-10~keV band is 
negligible in at least these radio galaxies. 
The weakness of the jet contribution can be understood in the framework 
of unified models if these galaxies are observed at intermediate inclination. 
In the case of
3C~120, an angle of 15~degrees between the jet axis and the line of sight
is sufficiently small to allow superluminal motion of the radio knots
to be observed 
(for bulk velocities $\gamma>8$) but sufficiently large that the Doppler 
factor is small,
$\delta < 3.0$, so that the intensity of beamed radiation is lower than 
that within the beam (i.e., at angles $< 1/\gamma < 7^\circ$) by a factor 
of 20  or more. 

What is unusual in 3C~120 is the width and intensity of the iron emission line
when the continuum is fitted with a single power law.

We mention two possible explanations for the iron line emission:
(1)
ASCA can not resolve a hard continuum component (like e.g. a Compton 
reflection component) and the iron line emission is overestimated. 
The OSSE observations do not resolve this ambiguity because
3C~120 was very weak in the 50-150~keV band during the 
simultaneous ASCA-OSSE 
observation, while OSSE data from literature probably represent
the source in a higher state of brightness.
In addition the reflection component peaks around 30-40~keV, i.e., in 
a spectral region not covered by our instruments, while the jet 
completely dominates the X-ray emission only above 100~keV (see Fig. 4).
In any case, the line remains rather broad and intense even if the reflection
component is included.

(2) Alternatively, the line is really broad. 
Martocchia $\&$ Matt (1996) have recently investigated 
the iron fluorescence line emitted by an accretion disc around 
a rotating (Kerr) black hole. 
They assume that the X-ray source is isotropic and located on the black hole
symmetry axis. Equivalent widths as large as 1-2~keV due to light
deflection and gravitational
blueshift of the primary radiation are expected if the 
distance, $h$, between the X-ray source and the disk 
is $h<6 r$, where $r=GM/c^2$ is the gravitational 
radius; i.e., $h<9\times 10^{13}$ cm
for M=$10^{8}M_{\odot}$. 
This explanation is not
implausible if relativistic jets are associated with the spin of the black hole
rather than with the angular momentum of the accreting matter. 
However, in this case the line profile is expected
to be not only very broad but also significantly redshifted, while the 
observed redshift, when the line is fitted with a gaussian, is 
not dramatic. 

As any iron line emission is expected
to be accompanied by a reflection continuum, 
the most obvious explanation is actually 
in term of a line plus
the reflection component. Unfortunately, the present data alone do not permit
to unambiguously disentangle the two components. 
We conclude that a Seyfert-like component dominates the X-ray emission of 
3C~120 at least in the 0.1-10~keV energy band.
To definitively exclude a jet contribution at higher energies 
requires additional hard X-ray observations.

\acknowledgments

We thank Tom Bridgman for providing OSSE data in XSPEC format and
for the very useful information on the OSSE archive.
PG and CMU acknowledge support from NASA grants NAG5-2538 and NAG8-1037.
RMS acknowledges financial support from a Research Associate NRC fellowship.

\newpage

\cl{\bf Figure Captions}

\bigskip

{\bf Figure 1.}
Light curves a) in the SIS0+SIS1 detectors (0.5-8 keV) and b) in the GIS2+GIS3
detectors (1-10 keV). The bin size is 500~sec. Significant variability
with amplitude $\sim30$\% is present on time scales of half a day.

\noindent
{\bf Figure 2.} Ratio of {\it a)} SIS and {\it b)}
GIS data to model for power-law fit.
An excess of emission is clearly evident at $\sim 6.2$~keV.

\noindent
{\bf Figure 3.} Iron line width versus iron line normalization at 
68$\%$, 90$\%$ and 99$\%$ confidence contour levels for a  power law
($\Gamma$ free to vary) plus gaussian profile model.  

\noindent{\bf Figure 4.}
Confidence contours in the two fit parameters, the amount of reflection (R) 
and the intrinsic width ($\sigma_{Fe}$) of the iron line,
for the GIS2+GIS3 data. Contours represent 68\%, 90\% and 99\% confidence 
levels. The strong anti-correlation between the two parameters is evident.

\noindent
{\bf Figure 5.}
ASCA--OSSE spectrum of 3C~120. 
The filled circle represents the contemporaneous February-March 1994 OSSE 
observations; the open circle is an average from published OSSE data. 
The OSSE data are best fitted by an extrapolation of the single ASCA power
law (which implies a broad iron line) or the reflection component. The
extrapolation of a second, flatter power law, which fits the 
non-contemporaneous high state OSSE data very well,
may indicate that the jet component is highly variable and is significant 
only at higher intensities.

\noindent
{\bf Figure 6.} 
ROSAT light curves for the observations of 10 September 1993. The bin size is 400 s.
A clear decrease in intensity, with doubling time $\sim 6$~hours, is seen.

\begin{center}
\footnotesize
\begin{tabular}{llcccl}
\multicolumn{6}{c}{{\bf Table 1:} Journal of Observations}\\
\hline
\multicolumn{1}{c}{Satellite}
&\multicolumn{1}{c}{Instrument}
&\multicolumn{1}{c}{Date}
&\multicolumn{1}{c}{Count Rate}
&\multicolumn{1}{c}{error}
&\multicolumn{1}{l}{F$_\nu^a$ }\\
& & &\multicolumn{1}{c}{(sec$^{-1}$)}&&($\mu$Jy) \\
\hline
&&&&&\\
ROSAT&PSPC& 1993 Mar 6 &1.740&0.030&10.76$^{+1.73}_{-1.30}$\\
&&&&&\\
     &    &1993 Aug 25 &3.010&0.038& 23.98$^{+4.12}_{-3.25}$ \\
&&&&&\\
     &    &1993 Sep 2 &2.040&0.030& 14.02$^{+2.19}_{-1.73}$ \\
&&&&&\\
     &    &1993 Sep 10 &2.413&0.035& 14.28$^{+2.46}_{-1.53}$\\
&&&&&\\
ASCA &SIS0&1994 Feb 17--19 &1.861&0.006&11.03$^{+0.25}_{-0.24}$ \\
&&&&&\\
     &SIS1&                  &1.412&0.006&\\
&&&&&\\
     &GIS2&                  &0.905&0.005&\\
&&&&&\\
     &GIS3&                  &1.095&0.006&\\
&&&&&\\
GRO  &OSSE$^b$&1994 Feb 17--Mar 1 &0.087& 0.31
& $1.32\pm 0.47\times10^{-3}$\\
&&&&&\\
     &        &1994 Mar 1--8      &     &     &  \\
&&&&&\\
\hline
&&&&&\\
\multicolumn{6}{l}{$^a$ - $\nu$=1 keV for ASCA and ROSAT; $\nu$=100 keV
for OSSE }\\
\multicolumn{6}{l}{$^b$ - Average count rate and flux density from both
the two OSSE observations}\\
\end{tabular}
\end{center}
\begin{center}
\footnotesize
\begin{tabular}{lcl}
\multicolumn{3}{c}{{\bf Table 2:} ASCA Time Variability} \\
\hline
\multicolumn{1}{c}{Instrument}
&\multicolumn{1}{c}{Bin size}
&\multicolumn{1}{r}{$\chi^2$/d.o.f}\\
&sec&\\
\hline 
&&\\
SIS0 & 100 & 376/180\\
     & 500 & 154/34\\
     &1000 &66/5\\
&&\\
SIS1 & 100 & 367/177\\
     & 500 & 176/33\\
     & 1000& 69/5\\
&&\\
GIS2 &100  & 490/176\\
     &500  & 122/49\\
     &1000 & 64/14\\
&&\\
GIS3& 100 & 294/174\\
    & 500 & 134/49\\
    & 1000& 69/14\\
&&\\
\hline
\end{tabular}
\end{center}

\begin{center}
\tiny
\begin{tabular}{lccccccccc}
\multicolumn{10}{c}{{\normalsize{\bf Table 3:} Fits to ASCA Spectrum of 3C 120
: 0.6-10 keV Energy Band}} \\
\hline
&&&&&&&&&\\
\multicolumn{1}{c}{Model$^a$}
&\multicolumn{1}{c}{$\Gamma_1$}
&\multicolumn{1}{c}{$N_H$}
&\multicolumn{1}{c}{$E_{Fe}$}
&\multicolumn{1}{c}{$\sigma_{Fe}$}
&\multicolumn{1}{c}{EW}
&\multicolumn{1}{c}{$\Gamma_2$}
&\multicolumn{1}{l}{$E_{\rm break}$}
&\multicolumn{1}{c}{R}
&\multicolumn{1}{c}{$\chi^2_r (d.o.f)$}\\
&&&&&&&&&\\
&
&($10^{21}$ cm$^{-2})$
&\multicolumn{1}{c}{(keV)}
&\multicolumn{1}{c}{(keV)}
&\multicolumn{1}{c}{(eV)}
&
&\multicolumn{1}{c}{(keV)}&
&\\
&&&&&&&&&\\
\hline
&&&&&&&&&\\
\multicolumn{10}{c}{\bf SIS0+SIS1 (0.6-5 keV)}\\ &&&&&&&&&\\
PL     & 2.00$\pm0.02$& 1.64$\pm0.07$ & & & & & && 1.11(66)\\
&&&&&&&&&\\
\hline 
&&&&&&&&&\\
\multicolumn{10}{c}{\bf SIS0+SIS1 (0.6-8 keV)}\\ &&&&&&&&&\\
PL     &1.96$^{+0.01}_{-0.02}$& 1.52$\pm0.06$& & & & & & &1.53(128)\\ 
&&&&&&&&&\\
PL+GA & 2.00& 1.63$^{+0.04}_{-0.03}$& 6.06$^{+0.22}_{-0.20}$& 
0.80$^{+0.25}_{-0.19}$ & 357$^{+111}_{-72}$& &&& 1.11(126)\\
&&&&&&&&&\\
PL+GA & 2.00$\pm0.02$ & 1.65$\pm0.07$ & 6.06$^{+0.26}_{-0.21}$
& 0.82$^{+0.28}_{-0.21}$& 399$^{+133}_{-122}$ & &&& 1.11 (125)\\
&&&&&&&&&\\
PL+DISKL$^b$& 2.00& 1.66$^{+0.03}_{-0.04}$& 6.2 &0.4 & 151$^{+56}_{-41}$
& &&&1.39(128)\\
&&&&&&&&&\\
PL+2GA&2.00 & 1.65$^{+0.04}_{-0.03}$&5.27$^{+0.08}_{-0.05}$ &0.1& 70
$^{+25}_{-36}$&& & & 1.34(125)\\
&&&6.20$\pm0.1$&0.1& 107$^{+35}_{-42}$&&&&\\
&&&&&&&&&\\
BK+GA & 2.00&1.64$^{+0.03}_{-0.04}$& 6.2& 0.4& 114$\pm64$&
1.79$\pm0.09$& 4.20$^{+0.33}_{-0.41}$&& 1.14(126)\\
&&&&&&&&&\\
BK+GA & 2.00& 1.64$^{+0.04}_{-0.03}$&6.2&0.4& 140$^{+50}_{-88}$& 
1.83$\pm0.07$&4.00&& 1.14(127)\\
&&&&&&&&&\\
BK+GA & 2.00& 1.63$\pm0.03$& 5.98$^{+0.31}_{-0.28}$& 
0.67$^{+0.36}_{-0.27}$ & 244$^{+311}_{-137}$ & 1.92$^{+0.21}_{-0.12}$ 
& 4.00 & & 1.11(125)\\
&&&&&&&&&\\
PLREF+GA & 2.00& 1.62$\pm0.04$& 6.2& 0.4& 171$^{+50}_{-86}$&
& &0.82$^{+0.47}_{-0.48}$ & 1.19(127)\\
&&&&&&&&&\\
PLREF + GA & 2.00& 1.63$^{+0.04}_{-0.03}$
& 6.06$^{+0.26}_{-0.22}$&0.80$^{+0.25}_{-0.21}$&372 $^{+94}_{-133}$& 
&&0.00$<0.53$& 1.12(125)\\
&&&&&&&&&\\
\hline 
&&&&&&&&&\\
\multicolumn {10}{c}{\bf GIS2+GIS3 (1-10 keV)}\\&&&&&&&&&\\
PL & 1.87$\pm0.03$ &0.73$^{+0.28}_{- 0.27}$& & & & & & & 
1.31(174)\\
&&&&&&&&&\\
PL+GA & 2.00& 1.53$\pm0.16$& 6.31$^{-0.19}_{+0.24}$&0.93$^{+0.25}_{-0.20}$&
674$^{+129}_{-144}$& & && 0.93(172)\\
&&&&&&&&&\\
PL+GA & 2.02$^{+0.08}_{-0.05}$& 1.68 $^{+0.49}_{-0.41}
$&6.30$^{+0.20}_{-0.19}$& 1.01$^{+0.44}_{-0.28}$&764$^{+456}_{-251}$
&& & &  0.93(171)\\
&&&&&&&&&\\
BK+GA& 2.00& 1.53$\pm0.16$ & 6.2 & 0.4 & 
222$^{+55}_{-82}$& 1.78$\pm0.08$& 3.99$^{+0.44}_{-0.50}$& & 0.92( 172)\\
&&&&&&&&&\\
BK+GA & 2.00& 1.52$<0.16$ & 6.24$^{+24}_{-20}$ & 
0.65$^{+0.01}_{-0.34}$ & 332$^{+229}_{-161}$& 1.84$^{+0.12}_{-0.10}$ 
& 4.00& & 0.91(171)\\
&&&&&&&&&\\
PLREF+GA & 2.00& 1.41$\pm0.15$& 6.2&0.4& 222$^{+56}_{-92}$& 
& &1.60$^{+0.61}_{-0.62}$ & 0.98(173)\\
&&&&&&&&&\\
PLREF+GA &2.00& 1.49$^{+0.19}_{-0.18}$& 6.30$\pm0.20$&
0.85$^{+0.30}_{-0.31}$& 
527$^{+240}_{-246}$& &&0.44$<1.52$& 0.93(171)\\
&&&&&&&&&\\
 \hline
\end{tabular}
\end{center}
\begin{center}
\tiny
\begin{tabular}{lccccccccc}
\multicolumn{10}{c}{{\normalsize{\bf Table 3:} - continued}} \\
\hline
&&&&&&&&&\\
\multicolumn{1}{c}{Model$^a$}
&\multicolumn{1}{c}{$\Gamma_1$}
&\multicolumn{1}{c}{$N_H$}
&\multicolumn{1}{c}{$E_{Fe}$}
&\multicolumn{1}{c}{$\sigma_{Fe}$}
&\multicolumn{1}{c}{EW}
&\multicolumn{1}{c}{$\Gamma_2$}
&\multicolumn{1}{l}{$E_{\rm break}$}
&\multicolumn{1}{c}{R}
&\multicolumn{1}{c}{$\chi^2_r (d.o.f)$}\\
&&&&&&&&&\\
&
&($10^{21}$ cm$^{-2})$
&\multicolumn{1}{c}{(keV)}
&\multicolumn{1}{c}{(keV)}
&\multicolumn{1}{c}{(eV)}
&
&\multicolumn{1}{c}{(keV)}&
&\\
&&&&&&&&\\
\hline
&&&&&&&&\\
\multicolumn {10}{c}{\bf SIS0+SIS1+GIS2+GIS3 (0.6-10 keV)}\\ 
&&&&&&&&&\\
PL+GA& 2.00& 1.62$^{+0.04}_{-0.03}$&6.23$^{+0.16}_{-0.15}$&
0.89$^{+0.17}_{-0.15}$& 509$^{+67}_{-85}$& && &1.02(302)\\
&&&&&&&&&\\
BK+GA & 2.00& 1.62$^{+0.04}_{-0.03}$&6.14$\pm0.18$& 0.69$^{+0.23}_{-0.24}$
&278$^{+159}_{-78}$  & 1.88$^{+0.10}_{-0.08}$& 4.00&& 1.012(301)\\
&&&&&&&&&\\
PLREF+GA& 2.00 &1.62$^{+0.04}_{-0.03}$& 6.23$^{+0.16}_{-0.15}$
& 0.89$^{+0.16}_{-0.15}$&
$497^{+77}_{-120}$  & &&0.00$<0.42$ &1.025(301)\\
&&&&&&&&&\\
\hline
&&&&&&&&&\\
\multicolumn{10}{l}{$^a$ PL=Power Law - GA=Gaussian line - 
BK=Broken Power Law - PLREF=Reflection Model - DISKL=
Gravitational shifted line profile}\\
\multicolumn{10}{l}{$^b$ The inclination of the disk is fixed to the
value i=15$^{\circ}$}\\
\end{tabular}
\end{center}
\clearpage

\footnotesize
\begin{center}
\begin{tabular}{lcccc}
\multicolumn{5}{c}{{\bf Table 4:} ROSAT Results: Power Law Model} \\
\hline
Date & & N$_H$ & $\Gamma$ & $\chi^2_r$(d.o.f) \\
&&&&\\
 & & (10$^{21}$ cm$^{-2}$) & &  \\\hline
& & & & \\
1993 Mar 6 && 2.26$^{+0.53}_{-0.43}$ & 2.54$^{+0.32}_{-0.28}$ &
       1.09(124)  \\
 & & & & \\
1993 Aug 25 && 3.12$^{+0.55}_{-0.50}$ & 3.27$^{+0.34}_{-0.31}$ &
       1.15(137)  \\
& & & & \\
1993 Sept 2 && 2.58$^{+0.50}_{-0.44}$ & 2.95$^{+0.31}_{-0.29}$ & 
       1.02(140)  \\
& & & & \\
1993 Sept 10 && 2.11$^{+0.54}_{-0.38}$ & 2.78$^{-0.34}_{-0.27}$ & 
       1.12(127)   \\
& & & & \\ \hline

\end{tabular}
\end{center}

\begin{center}
\begin{tabular}{lcccc}
\multicolumn{5}{c}{{\bf Table 5:} ROSAT Results: Power law + Raymond-Smith 
Model} \\
\hline
Date & & $\Gamma$ & kT &$\chi^2_r$(d.o.f) \\
& & & &\\
 & & & (keV)&  \\
\hline
& & & & \\
1993 Mar 6 &&1.55$^{+0.15}_{-0.18}$&  0.62$^{+0.18}_{-0.23}$ &
       1.13(123)  \\
 & & & & \\
1993 Aug 25 &&1.79$^{+0.15}_{-0.12}$ & 0.80$^{+0.08}_{-0.13}$ &
       1.32(136)  \\
& & & & \\
1993 Sept 2 && 1.79$^{+0.12}_{-0.12}$ & 0.68$^{+016}_{-0.13}$ & 
       1.08(139)  \\
& & & & \\
1993 Sept 10 && 1.96$^{+0.12}_{-0.13}$ & 0.67$^{+0.20}_{-0.24}$ & 
       1.17(126)   \\
& & & & \\ \hline

\end{tabular}
\end{center}

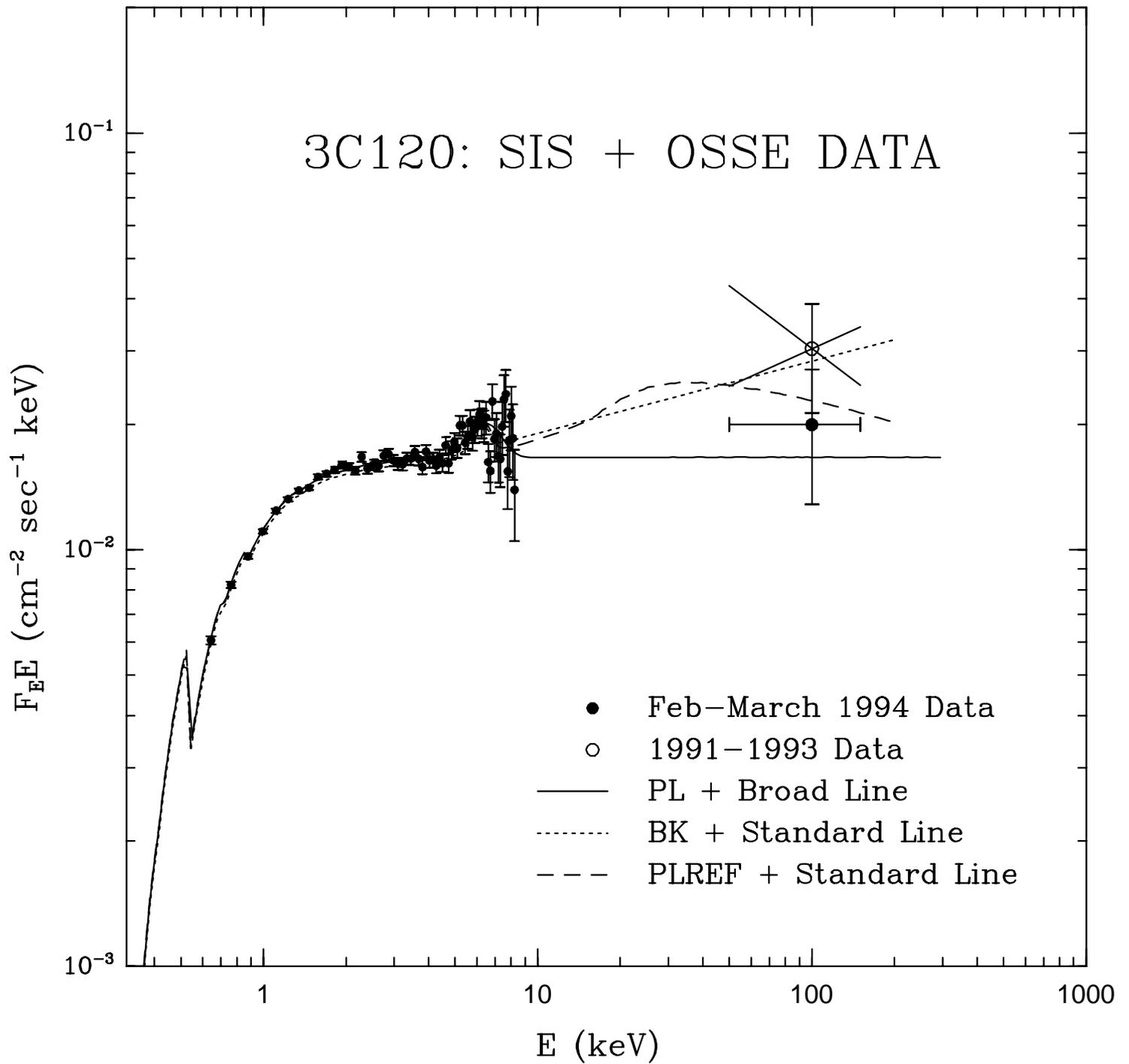
\end{document}